\documentstyle[emulateapj,onecolfloat]{article}

\newcommand{\Pla}{{{\bf P}^{\kappa}_{ij}}}

\newcommand{\tskip}{}
\newcommand{\fsky}{f_{\rm sky}}
\newcommand{\simlt}{\lesssim}
\newcommand{\simgt}{\gtrsim}

\submitted{Submitted \today}
\input epsf
\begin{document}
\twocolumn[
\title{Power Spectrum Tomography with Weak Lensing}
\author{Wayne Hu
}
\affil{Institute for Advanced Study, Princeton, NJ 08540}
\begin{abstract}
Upcoming weak lensing surveys on
wide fields will provide the opportunity to reconstruct the structure along
the line of sight tomographically by employing photometric redshift information
about the source distribution.  We define power-spectrum statistics,
including cross correlation between redshift bins, quantify the improvement
that 
redshift information can make in cosmological parameter estimation, and discuss
ways to optimize the redshift binning. 
We find that within the adiabatic cold dark matter
class of models, crude tomography using two or three redshift bins is 
sufficient to extract most of the information and improve the measurements
of cosmological parameters that determine the growth rate of structure by up 
to an order of magnitude.
\end{abstract}

]

\section{Introduction}

With new instruments such as MEGACAM at CFHT (\cite{Bouetal98}\ 1998) and the 
VST at the European Southern Observatory (\cite{Arnetal98}\ 1998), 
wide-field surveys detect the weak lensing of faint galaxies by large scale
structure will soon become a reality (see \cite{Mel98}\ 1998 for a 
recent review).  Weak lensing by large-scale structure produces
a correlated distortion in the ellipticities of the galaxies on
the percent level (\cite{Blaetal91} 1991; \cite{Mir91}\ 1991; 
\cite{Kai92} 1992) which can be used to measure
a two-dimensional projection of the intervening mass distribution 
(\cite{TysValWen90} 1990; \cite{KaiSqu93} 1993). 

If the redshift of the source galaxies are known, then more information
can be extracted out of weak lensing by tomography, i.e. differencing
the two-dimensional projected images to recover the three-dimensional
distribution.  In the absence of spectroscopy, approximate redshifts for
the faint galaxies can be determined through photometric techniques
(see e.g. \cite{Hogetal98} 1998 and references therein) and with the large
number of galaxies at $R \simlt 25$
($\sim 10^5$ deg$^{-2}$) the properties of
distribution can be known to good accuracy (\cite{Sel98} 1998).  
Indeed the weak lensing surveys
already plan to use photometric redshift information at least on a small
subsample to measure the low order
moments of the distribution such as its mean.  These are important for
determining the cosmological implications of the data (\cite{Smaetal95a} 1995;
\cite{ForMelDan96} 1995; \cite{LupKai97} 1997).  

The potential of tomographic techniques, especially in the wide-field
limit where the cosmological information is completely contained in 
the two-point functions or power spectra, remains largely unexplored.
Indeed most studies of weak lensing (e.g. 
\cite{JaiSel97} 1997;
\cite{Kai92} 1998;
\cite{HuTeg99} 1999) 
simply assume a delta-function distribution of galaxies
making tomography impossible.

In this {\it Letter}, we study the prospects for weak lensing tomography
within the framework of the adiabatic cold dark matter (CDM) class of models for
structure formation.  
We begin by defining the power spectrum statistics 
for an arbitrary set of galaxy redshift distributions.   
These are the power spectrum of the convergence map for each distribution
and the cross-correlation between the maps.  We then quantify
how much additional information can be extracted by subdividing a single
magnitude-limited sample into bins in redshift and analyzing
their joint power spectra and cross-correlation.  We conclude with
a discussion of how errors in photometric redshifts might affect
tomographic techniques.
 
\section{Power Spectra}

Generalizing the results of \cite{Kai98} (1992, 1998), we 
define the angular power spectra and cross-correlation of sky maps
of the convergence based on a series of galaxy redshift distributions
$n_i(z)$ (see also \cite{Sel98} 1998)
\begin{eqnarray}
\Pla(\ell)  &\equiv& {1 \over 2\ell+1} 
	\left< a^*_{(\ell m)\, i} a_{(\ell m) \, j} \right> \, \nonumber\\
	&\approx& 2\pi^2 \ell \int d D \, {{g_i(D) g_j(D)} \over 
	D_A^3(D)}
       \Delta_\Phi^2 (k_\ell,D) \,,
\label{eqn:ppsi}
\end{eqnarray}
where  $a_{(\ell m)\, i}$ are the spherical harmonic coefficients of
the maps.  Here
$D = \int_0^z (H_0/H) dz$ is the dimensionless comoving
distance and 
\begin{equation}
D_A(D)=
	\Omega_K^{-1/2} {\sinh(\Omega_K^{1/2}D)}
\end{equation}
is the dimensionless angular diameter distance, where
$\Omega_K=1 - \sum_i \Omega_i$ is the effective density in spatial curvature in 
units of the critical density.
The efficiency 
with which 
gravitational potential fluctuations $\Phi$, as measured by their dimensionless
power per logarithmic interval $\Delta_\Phi^2 \equiv k^3 P_\Phi/2\pi^2$, lens 
the given galaxy distribution $n_i$ is described by 
\begin{equation}
g_i(D) = D_A(D)
	\int_D^\infty dD'\, \left[ n_i { {dz \over dD} }\right](D') {D_A(D'-D) \over
	D_A(D')}  \,.
\end{equation}
Note that $n_i$ is normalized
so that $\int_0^\infty dz \, n_i(z)=1$.
Finally 
$k_\ell = \ell H_0 / D_A(D)$ 
	is the wavenumber which
	projects onto the angular scale $\ell$ at distance $D$.
For small fields of view, the spherical harmonics of order $\ell$ can be
replaced by Fourier modes with angular frequency $\omega$.

We use the \cite{PeaDod96} (1996) scaling relations to evalute
$\Delta_\Phi^2$ in the non-linear density regime.  Equation~(\ref{eqn:ppsi})
assumes that the redshift distributions are sufficiently wide to encompass
many wavelengths of the relevant fluctuations ($2\pi/k_\ell$) 
along the line of
sight so that the Limber equation holds even tomographically
(see \cite{Kai98} 1998).

These power spectra define the cosmic signal.  Shot noise in the
measurement from the intrinsic ellipticity of the galaxies adds white noise
to
the cosmic signal making the observed power spectra
\begin{equation}
{\bf C}_{ij}(\ell) = \Pla(\ell) + 
        {\left< \gamma_{\rm int}^2\right>\delta_{ij}/\bar n_i} \,, 
\label{eqn:covariance}
\end{equation}
where $\left< \gamma_{\rm int}^2 \right>^{1/2}$ is the rms intrinsic
shear in each component, and ${\bar n}_i$ is the number density of the galaxies
per steradian on the sky in the whole distribution $n_i(z)$.

\begin{figure}[t]
\centerline{\epsfxsize=3.5truein\epsffile{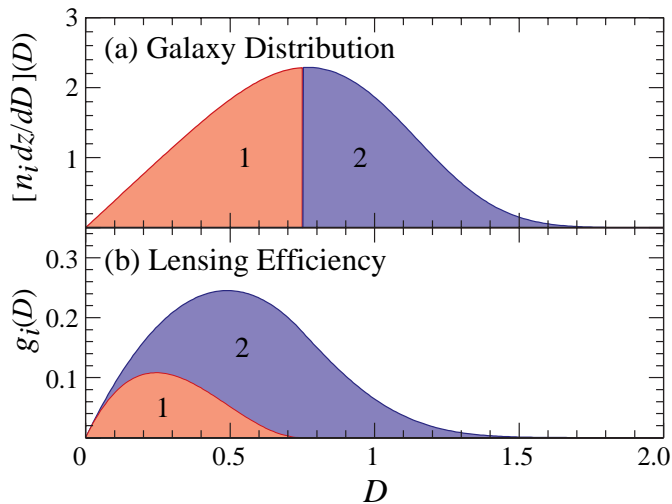}}
\caption{Subdividing the source population.  Partitioning the galaxies 
by the median redshift (or distance $D$) yields
lensing efficiencies with strong overlap.}
\label{fig:g}
\end{figure}

The distributions $n_i(z)$ need not be physically distinct galaxy populations.
Consider a total distribution $n(z)$ with
\begin{equation}
\left[ n {d z \over dD} \right](D) 
\propto D^\alpha \exp[-(D/D_*)^\beta] \,,
\label{eqn:distribution}
\end{equation}
which roughly approximates that of a magnitude-limited survey,
and take $\alpha=1, \beta=4$ for definiteness (assumed throughout
unless otherwise stated).  One can {\it subdivide}
 the sample into redshift bins to define the distributions $n_i(z)$.  
The power spectra for cruder partitions 
can always be constructed out of finer ones: if the 
$j$ and $k$ bins are combined, then 
\begin{eqnarray}
\bar n_{j+k}^2 {\bf P}_{(j+k)(j+k)}^\kappa &=& \bar n_{j}^2
{\bf P}_{jj}^\kappa + 2 \bar n_j \bar n_k {\bf P}_{jk}^\kappa
+ \bar n_k^2 {\bf P}_{kk}^\kappa \,, \nonumber\\
\bar n_{j+k} {\bf P}_{i(j+k)}^\kappa
&=& \bar n_j {\bf P}_{ij}^\kappa + \bar n_k {\bf P}_{ik}^\kappa \,.
\end{eqnarray}

In Fig.~\ref{fig:g}, we show an example where the galaxies
with $z< z_{\rm median}$ are binned into $n_1$ and the rest into $n_2$. 
Here and throughout we will take our fiducial
cosmology as an adiabatic CDM model with matter density $\Omega_m=0.35$, 
dimensionless Hubble constant $h=0.65$, baryon density $\Omega_b=0.05$, 
cosmological constant $\Omega_\Lambda=0.65$, neutrino mass $m_\nu=0.7$ eV, 
the initial potential power spectrum amplitude $A$, and tilt $n_S=1$.
 
We also plot in Fig. \ref{fig:g} the lensing efficiency
function $g_i(D)$.  Notice that despite having non-overlapping source
distributions (upper panel), the lensing efficiencies strongly
overlap (bottom panel) implying that the resulting convergence maps will have a
correspondingly large cross correlation.  This is of course because
the high and low redshift galaxies alike are lensed by low-redshift 
structures.   Also for this reason, there will be always be a stronger
signal in the high redshift bins.  This fact will be important for 
signal-to-noise considerations in choosing the bins.

\begin{figure}[t]
\centerline{\epsfxsize=3.5truein\epsffile{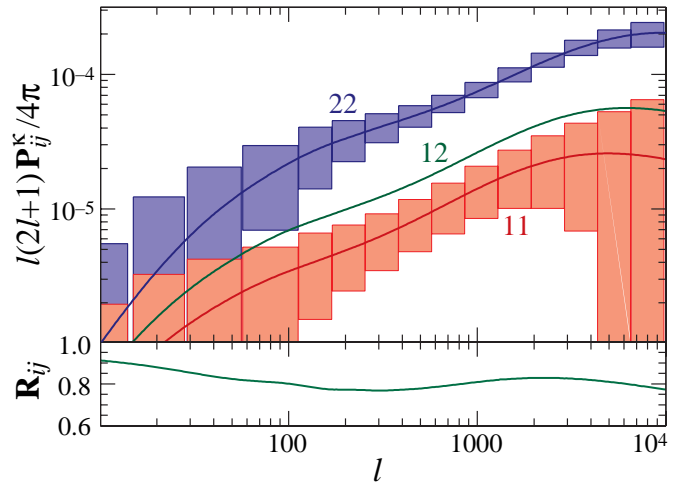}}
\caption{Power spectra and cross correlation for a subdivision in
two across the median redshift $z_{\rm median}=1$
and errors for a survey of $5^{\circ}$ on 
the side, $\left< \gamma_{\rm int}^2 \right>^{1/2}=0.4$, and
$\bar n = 2 \times 10^5$ deg$^{-2}$.
Note the strong correlation ${\bf R}_{ij}$ between the two power spectra make
the combination of the power spectra less constraining than a naive
interpretation of the individual errors would imply.}
\label{fig:power}
\end{figure}

All of these properties can be seen in Fig.~\ref{fig:power}, where we plot the
resultant power spectra and their cross correlation for
the equal binning of Fig.~\ref{fig:g}.

\section{Redshift Binning and Parameter Estimation}

While subdividing the sample into finer bins always increases the amount of
information, there are two considerations that limit the effectiveness of redshift
divisions.  The first is set by the shot noise from the intrinsic ellipticities
of the galaxies.  Once the number density $\bar n_i$ per bin
is so small
that shot noise surpasses the signal in equation (\ref{eqn:covariance}), further
subdivision no longer helps.  The point at which this occurs depends on the
angular scale of interest.  The greater number of galaxies encompassed by the
larger angular scales boosts the signal to noise (see Fig.~\ref{fig:power} and
\cite{Kai92} 1992).  Based on this criterion, one should separately subdivide the
data to extract the maximal large and small angle information. 

However there is a second consideration.  
If the lensing signal does not change significantly 
across the redshift range of the whole distribution, then subdivision
will not add information.  These considerations can be quantified by
considering the correlation coefficient between the power spectra of the subdivisions: 
${\bf R}_{ij} = {\bf P}^\kappa_{ij}/
({\bf P}^\kappa_{ii} {\bf P}^\kappa_{jj})^{1/2}$.  
For the model of Fig.~\ref{fig:power}, 
the power spectra are highly correlated (${\bf R}_{12} \sim 0.8$) even with only
two subdivisions.  Thus even though
there is enough signal to noise to subdivide the sample further, one gains little
information by doing so. 

One can combine these two considerations by diagonalizing the covariance
matrix and considering the signal to noise in the diagonal basis.  The appropriate
strategy for subdivision depends on the true redshift distribution of the 
galaxies and the model for structure formation.  One should therefore
perform this test on the actual data to decide how to subdivide the sample. 

Nevertheless, to make these considerations more concrete, let us consider the
specific goal of measuring the  
cosmological parameters $p_\alpha$ assuming that the underlying
adiabatic
CDM cosmology described above is correct.  
The Fisher information matrix can be used to quantify
the effect of subdivision.  It is defined as
\begin{equation}
{\bf F}_{\alpha\beta} = -\left< \partial^2 \ln L \over \partial p_\alpha \partial p_\beta 
\right>_ {\bf x} \,,
\end{equation}
where $L$ is the likelihood of observing a data set ${\bf x}$
given the true parameters
$p_1 \ldots p_\alpha$.

Generalizing the results of \cite{HuTeg99} (1998) to multiple
correlated power spectra, we obtain\footnote{When 
taking these derivatives the redshift distribution $n_i(z)$ is held fixed
as opposed to the distance distribution $[n_i(z)dz/dD]$ in 
equation~(\ref{eqn:distribution}).}
\begin{equation}
{\bf F}_{\alpha\beta} = \sum_{\ell=2}^{\ell_{\rm max}} 
		({\ell+1/2}) \fsky {\rm tr} [{\bf C}^{-1} {\bf C}_{,\alpha} 
		{\bf C}^{-1} {\bf C}_{,\beta}]\,,
\label{eqn:Fisher}
\end{equation}
under the assumption of Gaussian signal and noise,
where $\fsky$ is fraction of sky covered by the survey, the covariance matrix
${\bf C}$ was defined in equation~(\ref{eqn:covariance}), and commas denote partial
derivatives with respect to the cosmological parameters $p_\alpha$.  
We take $\ell_{\rm max}=3000$ to approximate the increased covariance
due to the nonlinearities producing non-Gaussianity in the signal
(\cite{ScoZalHui99} 1999).
Since the variance of an unbiased estimator of a
parameter $p_\alpha$ cannot be less than $\sigma(p_\alpha)=({\bf F}^{-1})_{\alpha\alpha}$, 
the Fisher matrix quantifies the best statistical errors on parameters possible with 
a given data set.  

For the purposes of this work, the 
absolute errors on parameters are less relevant than the 
improvement in errors from subdividing the data (see
\cite{HuTeg99} for the former).  
We therefore test a 4 dimensional subset of the adiabatic CDM parameter space
to see how subdivision
can help separate initial power ($A$) 
from the various contributors to the redshift-dependent evolution of power
($\Omega_\Lambda$,$\Omega_K$,$m_\nu$). 
For reference the standard errors $\sigma_\alpha$
for this parameter space without subdivision
are given in Table 1.
Errors in the full parameter space would be increased but note that the neglected
parameters $(\Omega_m h^2$,$\Omega_b h^2$, and $n_S$) are exactly those that 
the CMB satellite experiments should constrain precisely (see e.g. 
\cite{Junetal96} 1996; \cite{EisHuTeg99} 1999).

As an example, we take a sample with $z_{\rm median}=1$ and
$\bar n = 2 \times 10^5$ deg$^{-2}$ 
as appropriate for a magnitude limit of $R \sim 25$ (see 
\cite{Smaetal95b} 1995b). 
The signal to noise in the full sample is quite high, e.g.\ at
$\ell=100$, $S/N \sim 25$.  Thus we expect that subdividing the sample
should improve parameter estimation.

\begin{center} 
{TABLE 1\\[4pt] \scshape Parameter Estimation for $z_{\rm median}=1$} \\[3pt]
\begin{tabular}{llllllll}
\tskip\tableline\tableline\tskip
\tskip\tskip  $p_\alpha$ & $\sigma_\alpha \fsky^{1/2}$
& \multicolumn{6}{c}{Error Improvement} \\
& 1 &
2(${1 \over 2}$) & 2(${1 \over 4}$) & 2(${1 \over 8}$) & 3(${1 \over 3}$) 
& 3(${1 \over 4}$) & 3(${1 \over 8}$) \vphantom{\Big[}\\
\tskip\tableline\tskip\tskip
$\Omega_\Lambda$
	&	0.040
        &       6.5
        &       6.9
	& 	5.7
        &       7.2
	&	7.7
	&	6.9\\
$\Omega_K$
	&	0.023
        &       2.9
        &       3.1
	&	2.9
        &       3.3
	&	3.5
	&	3.2\\
$m_\nu$
	& 	0.044
        &       1.7 
        &       2.0
	& 	2.1
        &       2.1
	&	2.2
	&	2.2\\
$\ln A$
	&	0.064
        &       1.7
        &       2.0
	&	2.0
        &       2.1
	&	2.2
	&	2.1
\end{tabular}
\end{center}

\begin{figure}[t]
\centerline{\epsfxsize=3.5truein\epsffile{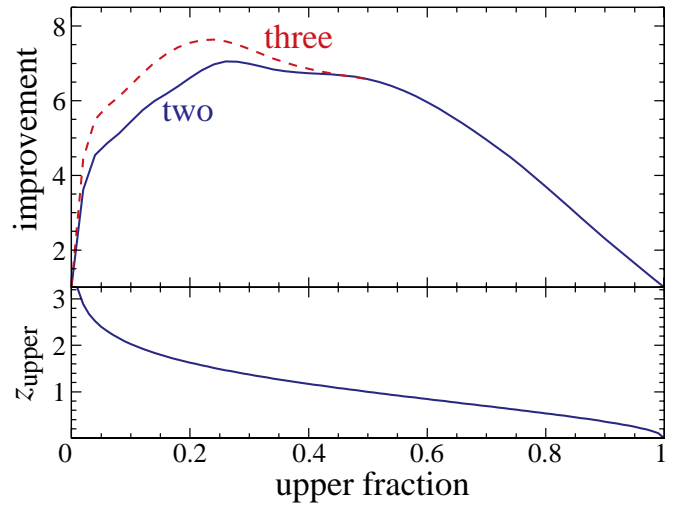}}
\caption{Tomographic error improvements on $\Omega_\Lambda$ for
$z_{\rm median}=1$.  Upper panel: improvement as 
a function of the fraction of galaxies in the upper redshift bin for 2 
bins versus 3 bins (same fraction in upper two bins).  
Lower panel: redshift corresponding to the upper division.}
\label{fig:improvement}
\end{figure}

As shown in Table~1, subdividing 
this sample in equal halves, denoted as 2(1/2),
improves the errors $\sigma_\alpha$ by a factor of 2 to 7.   Since the signal in the lower
redshift bin is smaller than in the higher redshift bin, it suffers comparatively
more from the intrinsic noise variance.  One can optimize the binning to correct
for this effect.  
Dividing
the sample so as to isolate the upper quarter [2(1/4)] improves
the errors modestly whereas isolating the upper eighth deproves them.
We plot the full range as a function of the fraction of galaxies
in the upper bin in
Fig.~\ref{fig:improvement}.  Notice that though the improvement factor
is roughly flat from $0.15-0.5$, it drops rapidly 
when noise dominates either the upper or lower fraction.  If the signal
were the same in both bins, this would occur at $0.04$ and $0.96$ for
$\ell=100$.  The fact that the true improvement is skewed to smaller
upper fractions reflects the fact that the signal increases to higher
redshifts.

Moving to three divisions
makes only a small improvement over two.  In Table~1 we give the results
of taking 3 bins with an equal number of galaxies in the upper two bins,
e.g.\ [3(1/4)] represents a division by number of ($1/2$, $1/4$, $1/4$).
In fact the errors for three bins can be higher than those with two
if not chosen wisely.

We conclude that for a redshift distribution 
of the form given by equation~(\ref{eqn:distribution})  
with $z_{\rm median}=1$, $\alpha=1$ and 
$\beta=4$, 
crude partitioning suffices to
regain most of the redshift information
in adiabatic CDM models where the change in
the growth rate across the distribution is slow and controlled by a small
number of cosmological parameters 

How robust are these conclusions against changes in the distribution 
and model?  A wider redshift distribution offers greater opportunities for
tomography.
For example, let us widen the distribution by taking
$\beta=2$ 
in equation~(\ref{eqn:distribution}).  Then the gains by simply halving the distribution
are a factor of 9.7 for $\Omega_\Lambda$; going to a 3(1/4) scheme raises
this to 12.  

These considerations are also relevant for
deeper surveys.  
Consider a survey with $z_{\rm median}=2$ and $\bar n=3.6 \times 10^5$
deg$^{-2}$.
The parameter estimation results are given in Table 2.  Not only is the overall
improvement from subdivision larger (up to a factor of 24 for three bins) 
but the relative improvements
between parameters also changes.    This is because even within the adiabatic
CDM paradigm the importance of the different parameters in determining the growth
of structure depends on redshift.

Perhaps the most important aspect of weak lensing tomography is that
it has the ability to falsify the underlying adiabatic CDM model.
For this reason, it is wise to examine the power spectra from the redshift
bins directly, since these are the observables, rather than jump directly
to modelling the data with parameters under the adiabatic CDM framework.
For example, tomography may show that the component that accelerates 
the expansion of the universe is not the cosmological constant 
or call into question the
fundamental assumption 
that structure forms through the gravitational instability of cold 
dark matter.


\begin{center}
{TABLE 2\\[4pt] \scshape Parameter Estimation for $z_{\rm median}=2$}\\[3pt]
\begin{tabular}{lllllll}
\tskip\tableline\tableline\tskip
\tskip\tskip  $p_\alpha$ & $\sigma_\alpha \fsky^{1/2}$ 
& \multicolumn{5}{c}{Error Improvement} \\
& 1 &  2(${1 \over 2}$) & 2(${1 \over 4}$) & 
2(${1 \over 8}$) & 
3(${1\over 4}$) &
3(${1\over 8}$)  \vphantom{\Big[}
\\
\tskip\tskip\tableline\tskip\tskip
$\Omega_\Lambda$
	& 	0.063
        &     19 
        &     21  
	&     20
	&     24	
	&     24\\
$\Omega_K$
	&	0.030
        &     6.7 
        &     7.7 
	&     8.0
	&     8.9	
	&     9.1\\
$m_\nu$
	&	0.027
        &     2.3  
        &     2.9  
	&     3.0	
	&     3.2
	&     3.4	\\
$\ln A$
	&	0.040
        &     2.1 
        &     2.6 
	&     2.1
	&     3.1
	&     3.2
\end{tabular}
\end{center}

\section{Discussion}

We have shown the precision with which cosmological parameters can
be measured from a weak-lensing survey can be significantly enhanced
by tomographically determining the evolution of the
statistical properties of 
large-scale structure across the finite redshift width of the 
source distribution.  Crude redshift binning of the data can recover
most of the statistical information contained in the redshifts.  
For example, most of the gain for a magnitude limited survey
with $z_{\rm median}=1$, under the adiabatic cold dark matter paradigm, 
comes from separating out
the upper and lower redshift halves of the distribution. 
For wider distributions and stronger rates of change
in the growth of structure, more information can be
extracted by finer binning, especially of the higher redshift portion of the
sample where the signal is greater.
The appropriate number of bins can be empirically determined by 
examining the correlation between bins and the noise properties of
the data.

We have been assuming that the individual redshifts of the galaxies
will be known sufficiently precisely to determine the redshift distribution 
of the subsamples.  Realistically, the redshift information will be limited
by the accuracy of photometric redshift techniques which currently
show errors of $\Delta z \sim 0.1$ (68\% CL) for $0.4 \lesssim z \lesssim
1.4$ (\cite{Hogetal98} 1998).  
While statistical errors on the large samples of galaxies considered
above are negligible, systematic errors or biases in the technique may
cause problems.  It is beyond the scope of this letter to test these
issues fully.  To give some feel for their effect, let us consider
the median redshift $z_{\rm median}$ as an additional parameter with
a prior uncertainty from photometric redshifts of the full individtual 
error $0.1$.  
Including this uncertainty degrades
the precision in the parameters by $3\%$ in the worst case.

While this effect is negligible, more worrying is a bias that is a function
of redshift, especially in the largely untested regime $1.4 \lesssim z 
\lesssim 2$, as that can shift the difference between the power spectra
of the subdivisions.  Isolating the few percent of galaxies at $z \simgt
2.5$, where the techniques are tested, yields gains that are comparable to
the optimal division (see Fig.~\ref{fig:improvement} lower panel), but the
compactness of such galaxies poses an obstacle for measuring the
lensing distortion from the ground (\cite{Steetal96} 1996).  
Despite these caveats, this study shows that tomography with weak lensing is
both possible and would substantially improve the precision with which
we can measure the growth of structure in the universe. 

{\it Acknowledgements:} 
I would like to thank D.J. Eisenstein, D.W. Hogg, J. Miralda-Escude, 
D.N. Spergel, M. Tegmark,
J.A. Tyson, M. White, and D. Wittman for useful conversations. 
W.H.\ is supported by the Keck Foundation, a Sloan Fellowship,
and NSF-9513835

\end{document}